\documentstyle[aps,preprint,epsfig]{revtex}

\def\be{\begin{equation}}
\def\ee{\end{equation}}
\def\ba{\begin{eqnarray}}
\def\ea{\end{eqnarray}}

\def\ga{\mathrel{\mathpalette\fun >}}
\def\fun#1#2{\lower3.6pt\vbox{\baselineskip0pt\lineskip.9pt
        \ialign{$\mathsurround=0pt#1\hfill##\hfil$\crcr#2\crcr\sim\crcr}}}

\begin{document}

\draft

\title{Probing the Evolution of the Dark Energy Density\\
with Future Supernova Surveys}
\author{Yun Wang, Veselin Kostov}
\address{Department of Physics \& Astronomy,
University of Oklahoma, Norman, OK 73019}
 
\author{Katherine Freese}
\address{Michigan Center for Theoretical Physics,
Physics Department, University of Michigan,\\
Ann Arbor, MI 48109}

\author{Joshua A. Frieman}
\address{NASA/Fermilab Astrophysics Center\\
Fermi National Accelerator Laboratory\\
Batavia, IL 60510\\
and\\
Department of Astronomy and Astrophysics and\\
Center for Cosmological Physics\\
The University of Chicago\\
5640 South Ellis Avenue, Chicago, IL 60637}

\author{Paolo Gondolo} 
\address{Department of Physics, 
University of Utah,
115 South 1400 East, Suite 201, 
Salt Lake City, UT 84112-0830}

\date{Februrary 2, 2004}
\maketitle

\begin{abstract}
  
  The time dependence of the dark energy density can be an important
  clue to the nature of dark energy in the universe.  We show that
  future supernova data from dedicated telescopes (such as SNAP), when
  combined with data of nearby supernovae, can be used to determine
  how the dark energy density $\rho_X(z)$ depends on redshift, if
  $\rho_X(z)$ is not too close to a constant.  For quantitative
  comparison, we have done an extensive study of a number of dark
  energy models.
  Based on these models we have simulated data sets in order to show
  that we can indeed reconstruct the correct sign of the time
  dependence of the dark energy density, outside of a degeneracy
  region centered on $1+w_0 = -w_1 z_{max}/3$ (where $z_{max}$ is the
  maximum redshift of the survey, e.g., $z_{max}=1.7$ for SNAP).  We
  emphasize that, given the same data, one can obtain much more
  information about the dark energy density directly (and its time
  dependence) than about its equation of state.

\end{abstract}


\newpage

\narrowtext

\section{Introduction}

Most of the energy in our universe is of unknown nature to us.  The
amount of this dark energy has been determined by recent experiments,
including the Wilkinson Microwave Anisotropy Probe (WMAP) satellite
observations \cite{wmap} of the anisotropy in the cosmic microwave
background radiation.  Our universe is spatially flat (the
three-dimensional equivalent of a two-dimensional plane), with roughly
27\% matter and 73\% dark energy.  Determining the nature of this dark
energy is one of the major fundamental challenges in astronomy and
physics today.

There are many plausible candidates for dark energy.  For example, (1)
a cosmological constant, i.e., constant vacuum energy
originally proposed by Einstein 
in his equations of general relativity, (2) a time
dependent vacuum energy, or scalar field known as ``quintessence'',
that evolves dynamically with time\cite{yaya}, or (3) modified
Friedmann equation, e.g. the Cardassian models
\cite{card,mpcard,card03}, that could result as a consequence of our
observable universe living as a 3-dimensional brane in a higher
dimensional universe.  Other proposed modifications to the Friedmann
equation include \cite{others}.  The time-dependence of the density of
dark energy can reveal the nature of dark energy at a fundamental
level.

A powerful probe of dark energy is type Ia supernovae (SNe Ia), which
can be used as cosmological standard candles to measure how distance
depends on redshift in our universe. Observations of SNe Ia have
revealed the existence of dark energy in the universe \cite{SN1,SN2}.
Current SN Ia data are not yet very constraining on the nature of the
dark energy \cite{Wang04}.

The distance-redshift relation of observed supernovae depends on the
nature of dark energy.  Most researchers have chosen to parametrize
dark energy by its equation of state parameter. However, it has been
shown \cite{mbs,barger} that it is extremely difficult to constrain the
time-dependence of the dark energy equation of state using supernova
searches (or any other technique relying on the luminosity distance);
hence one might worry that one cannot differentiate between different
dark energy models.  Fortunately, it has been shown that one can do
much better if one parametrizes the dark energy by its density
directly, instead of its equation of state 
\cite{Wang01a,Wang01b,Tegmark02,Daly03}.
The dark energy density is a more fundamental parameter than the dark
energy equation of state parameter.  Obtaining the equation of state
parameter requires one to perform an additional integral (compared to
obtaining the dark energy density); this integral smears out much of
the information one could otherwise learn.  Hence, given the same
data, the uncertainties of the constraints on the dark energy density
should be {\it smaller} than that of the constraints on the dark
energy equation of state.

In this paper we focus on extracting information about the dark energy
density directly.  We will show how well, using future supernova data,
one can determine whether the dark energy density changes with time,
and whether it increases or decreases with time.

We begin in Section II with the basic equations for using supernovae
to study dark energy.  In Section III we present four theoretical
models which we will study to see how well we can reconstruct the time
dependence of the dark energy: Models 1, 2, and 3 have dark energy
density that is constant, increasing, and decreasing in time
respectively. As our fourth set of models we consider those
parametrized by an equation of state $w_X(z)=w_0+w_1 z$.  In Section
IV, we simulated SNIa data for these models.  In Section V, we use the
adaptive iteration method to see how well we can reconstruct the time
dependence of the dark energy density for these models: we use three
test functions with different time dependences to see which one best
matches the data.  For each model we then run 1000 Monte Carlo samples
to obtain error bars for our fit.  The results are presented in
Section VI, followed by the conclusions.

\section{Basic Equations}

Type Ia supernovae (SNe Ia) are our best candidates for cosmological
standard candles, because they can be calibrated to have small
scatters in their peak luminosity \cite{Phillips93,Riess95}.  The
distance modulus for a standard candle at redshift $z$ is \be
\label{eq:mu0p}
\mu_p (z)\equiv m-M= 5\,\log\left( \frac{ d_L(z)}{\mbox{Mpc}} \right)+25,
\ee
where $m$ and $M$ are the apparent and absolute magnitudes of the standard
candle, and $d_L(z)$ is its luminosity distance. 

In a flat Friedmann-Robertson-Walker universe (which we assume in this
paper since it is strongly suggested by current CMB data \cite{omega1,wmap})
the luminosity distance $d_L(z)$ is given by
\cite{Weinberg72} \be
\label{eq:r(z)}
d_L(z)= \frac{(1+z) c}{H_0} \, \int_0^z\frac{dz'}{E(z')} , \ee where
$H_0$ is the present value of the Hubble constant, and \be E(z) \equiv
\sqrt{ \Omega_m(1+z)^3+ \Omega_X f_X(z) }.  \ee Here $\Omega_m$ is the
present value of the matter density in unit of the critical density
$\rho_{\rm crit} = 3 H_0^2/(8\pi G)$, and $\Omega_X$ is the present
value of the dark energy density in the same units.  

The condition for a flat universe imposes the relation 
\be
\Omega_m + \Omega_X = 1.
\ee

The dark energy density function
\be
f_X(z) = \frac{ \rho_X(z) } {\rho_X(0)}
\ee
describes the redshift dependence of the dark energy density $\rho_X(z)$.

We note that Cardassian models \cite{card,mpcard,card03} contain matter and
radiation only (no vacuum), so that in those models, $\Omega_m$ and
$\Omega_X$ are used to refer to {\it effective} observed matter
density and dark energy density respectively.

For a given cosmological model with dark energy density $\rho_X(z)$,
or dark energy function $f_X(z)$, we can compare the measured distance
modulus of SNe Ia at various redshifts with the predicted distance
modulus of a standard candle at these redshifts.  A systematic
comparison spanning all plausible models yields constraints on the
dark energy density $\rho_X(z)$.

\section{Dark Energy Density Functions}
\label{sec:2}

In this paper, we are interested in finding what information future SN
Ia data can give us about the redshift dependence of the dark energy
density. For this purpose, we consider four classes of dark energy
densities: (1) constant with redshift, (2) increasing with redshift,
(3) decreasing with redshift, and (4) a grid of models which includes
some that are non-monotonic with respect to redshift.  For each of
these four classes we choose simple representative models.  Some of
the models we choose can be parametrized by a simple equation of
state, \be w_X(z)=w_0+w_1 z \,
\label{eq:w(z)}
\ee in order to allow simple comparison with previous work in the
literature.  The simple parametrization of Eq.(\ref{eq:w(z)}) also
allows us to estimate how small a deviation from a constant dark
energy density can be determined by our technique.  The dark energy density is
constant only for $w_0=-1$ and $w_1=0$; any other values of $w_0$ or
of $w_1$ parametrize deviations from a constant.  However, we stress
that it is the dark energy density itself, $\rho_X(z)$, that we
extract from simulated data, as it is the more fundamental and more
easily extracted quantity.

\vskip 0.2in
\centerline{\bf Four Models}
\vskip 0.1in

The four sample theoretical models we consider are (see Table 1):

\begin{center}
Table 1\\
{\footnotesize{Dark Energy Models}}

{\footnotesize
\begin{tabular}{|l|l|l||}
\hline 
Model & model parameters & $\rho_X'(z)$ \\  
\hline 
Model 1: $\Lambda$CDM    &  $\Omega_m=0.3$, $\Omega_{\Lambda}=0.7$   & 
$\rho_X'(z)=0 $\\
Model 2: quintessence model  & $\Omega_m=0.3$, $w_q(z)=-1+0.5\,z$   & 
$\rho_X'(z)\geq 0 $ \\
Model 3: MP Cardassian model  & $\Omega_m^{obs}=0.3$, $n=0.2$, $q=2$ &
$\rho_X'(z) \leq 0 $ \\
Model 4: grid of models & $w_X(z)=w_0+w_1 z$ with {\rm arbitrary} $w_0, w_1$ & 
{\rm all} $\rho_X'(z)$ \\
\hline 
\end{tabular}
}
\end{center}

(1) MODEL 1: For a constant dark energy density,  $\rho_X'(z) = 0$, we
have a cosmological constant model ($w_0=-1$ and $w_1=0$).  

(2) MODEL 2: For an increasing dark energy density, $\rho_X'(z) \geq 0$,
we choose a quintessence model with equation of state
$w_X(z)=-1+0.5z$.  Popular quintessence models
in which the quintessence field tracks the matter field have
$\rho_X'(z) \geq  0$ \cite{barger}.

(3) MODEL 3: For a decreasing dark energy density,  $\rho_X'(z) \leq 0$,
we choose a Modified Polytropic (MP) Cardassian model \cite{mpcard}
with $n=0.2$ and $q=2$,
 \begin{equation}
 \rho_X(z) = \rho_{\rm crit} \Omega_m (1+z)^3 \left\{ \left[ 1 +
   \frac{\Omega_m^{-q}-1}{(1+z)^{3q(1-n)}} \right]^{
   \frac{1}{q} } - 1 \right\}.
 \end{equation}
 
 MP Cardassian models can have either $\rho_X'(z)\geq 0$ or
 $\rho_X'(z) \leq 0$.  Our previous paper \cite{card03} shows the
 regions of parameter space that fall into the two regimes. We also
 discussed there that MP Cardassian models can be found with $w_X
 =p_X/\rho_X< -1$ but with $w=p/\rho \geq -1$, so that the dominant
 energy condition holds (here, $w$ refers to the total energy density
 whereas $w_X$ refers only to the new component in the Friedmann
 equation that mimics a dark energy).  An effective $w_X < -1$ is
 consistent with recent CMB and large scale structure data
 \cite{sch,melchiorri}.

(4) MODEL 4: We consider a grid of models of the form
Eq.(\ref{eq:w(z)}), for $ -1.2 \leq w_0 \leq -0.5$, and $-1.5 \leq w_1
\leq 0.5$, and a grid spacing of $\Delta w_0=0.1$ (0.2 for $w_0\leq
-1$) and $\Delta w_1=0.1$.  This represents a total of 147 models.
This grid includes models with $\rho_X'(z)$ that is monotonically
increasing or decreasing with redshift, as well as models with dark
energy density that is non-monotonic with redshift.  
In a moment we will show which values of $w_0$ and $w_1$ 
correspond to a non-monotonic dark energy density.

Although models of the form of Eq.(\ref{eq:w(z)}) do not correspond
exactly to physically motivated models, it is interesting to note that
they can approximate a wide range of models. For example, Model 3 (MP
Cardassian model with $n=0.2$, $q=2$, and $\Omega_m =0.3$) can be
roughly approximated by a dark energy model with $w_X(z)=-1.10-0.35
z$.  The approximate equivalence of these models does not extend to
the behavior of dark energy fluctuations but is limited to the average
properties of the energy density.

The dark energy density corresponding to Eq.(\ref{eq:w(z)}) is
\cite{Wang01a} \be \rho_X(z) = \rho_X(0) e^{3w_1 z}\,
(1+z)^{3(1+w_0-w_1)}.  \ee Its derivative with respect to redshift
follows as 
\begin{equation} 
\label{eq:rhoprime}
\rho_X'(z) = \rho_X(z) \frac{3(1+w_0+w_1 z)}{1+z}.  
\end{equation}
Thus $\rho_X'(z)$ has the same sign as $ 1+w_0+w_1 z $. 
\bigskip

{\it Nonmononotic models:} Since $z$ must be positive, models with (i)
$w_0<-1$ and $w_1>0$, and models with (ii) $w_0>-1$ and $w_1<0$ have
non-monotonic dark energy density $\rho_X(z)$. Models of type (i) have
$\rho'(z)<0$ for $z<z_{crit}$ and $\rho'(z)>0$ for $z> z_{crit}$ where
\be
\label{eq:zcrit}
z_{crit} = \left|(1+w_0)/w_1 \right |.
\ee
Models of type (ii) have $\rho'(z)>0$ for $z<z_{crit}$ and
$\rho'(z)<0$ for $z> z_{crit}$.

All models other than types (i) and (ii) above have monotonic dark
energy densities, decreasing with redshift for $w_0<-1$ and $w_1<0$
and increasing with redshift for $w_0>-1$ and $w_1>0$.

\section{Simulated data}

We now construct simulated SN Ia data for dark energy Models 1, 2, 3, and
4 defined above, and investigate if we can recover the original theory
from the simulated data.  

We simulate the data by distributing SNe Ia in $z$ randomly per 0.1
redshift interval, with the total number per redshift interval as
expected for SNAP.  Here we assume that SNAP will obtain all SNe Ia in
its survey fields up to $z=1.7$ \cite{Tarle02}, similar to a supernova
pencil beam survey \cite{Wang00a,Wang01b}.  We increase the number of
SNe Ia at low redshifts, such that there are a minimum of 50 SNe Ia
per 0.1 redshift interval at $z\leq 0.5$. We assume that these
additional low redshift supernovae will come from surveys of nearby
SNe Ia. Thus each simulated data set consists of 2300 SNe Ia.

The measured distance modulus for the $l$-th SN Ia is
\be
\mu_0^{(l)}= \mu_p^{(l)}+\epsilon^{(l)}
\label{mu_0}
\ee
where $\mu_p^{(l)} = \mu_p(z_l)$ is the theoretical
prediction in our dark energy model for a SN Ia at redshift $z_l$ [see Eq.(\ref{eq:mu0p})], and 
$\epsilon^{(l)}$ is the uncertainty in
the measurement, including observational errors and intrinsic scatters in the
SN Ia absolute magnitudes.  
In the simulated data set, we 
draw the dispersion $\epsilon^{(l)}$ for the $l$-th SN Ia from
a Gaussian distribution with variance $\Delta m_{int}=0.16\,$mag. 
We simulate one set of data for each of the four models 
described in the Table.

\section{Estimation of Dark Energy Functions}
\label{sec:5}

We recover the dark energy function from each simulated data set.
A number of techniques may be used to achieve this reconstruction.
In this paper we use
the adaptive iteration method \footnote{For a complementary method,
see \cite{Wang04} which uses a Markov Chain Monte Carlo (MCMC) technique.}
introduced by Wang \& Garnavich
(2001)\cite{Wang01a} and Wang \& Lovelace (2001)\cite{Wang01b} and
described briefly here.  Our current study builds on our previous work
\cite{card03}.  There we proposed a technique for determining the
correct sign of $\rho_X'(z)$ if $\rho_X(z)$ is not too close to a
constant. To quantify how close to a constant we can go, in this paper
we perform Monte Carlo simulations in order to obtain error bars for
our results.

\subsection{Adaptive Iteration Method}

We start from the simulated data sets constructed from each of the
four models described above.  Given our data set, we now proceed as
though we did not know which model it came from.  We pretend we know
nothing about the form of $\rho_X(z)$.  In attempting to reconstruct
the dark energy density, we run through a series of test functions: we
allow the test function $\rho_X^{\rm test}(z)$ to be an arbitrary
function.  To approximate the function, we parametrize it by its value
at $N+1$ equally spaced redshift values $z_i$ ($i=0,1,2,\ldots,N$,
$z_0=0$, $z_{N}=z_{max}$).  The values of $\rho_X^{\rm test}(z)$ at
other redshifts are given by linear interpolation, i.e.,
\ba
\label{eq:f(z)}
&&\rho_X^{\rm test}(z)=\left( \frac{z_i-z}{z_i-z_{i-1}} \right)\, \rho_{i-1}+
\left( \frac{z-z_{i-1}}{z_i-z_{i-1}} \right)\, \rho_i,
 \hskip 1cm z_{i-1} < z \leq z_i, \nonumber \\
&& z_0=0, \,\, z_{N}=z_{max} .
\ea
The values of the dark energy density
$\rho_i$ ($i=1,2,...,N)$ are the independent variables
to be estimated from data.  Again, we proceed as though we
had absolutely no information on the function $\rho_X(z)$,
and treat it as a completely arbitrary function.

It is convenient to trade the $N+1$ parameters $\rho_i$ with the $N$
parameters $f_i$ and the single parameter $\Omega_m$, where 
\begin{equation} 
\qquad
f_i = \frac{\rho_i}{\rho_0} \quad (i=1,2,\ldots,N) \,\,\,\,\,\,\, {\rm
  and} \,\,\,\,\,\,\, \Omega_m = 1- \frac{\rho_0}{\rho_{\rm crit}} .
\end{equation}
We define 
\begin{equation} 
\rho_0 \equiv \rho_X(0) 
\end{equation} 
and take $\rho_{\rm
  crit} \equiv 3 H_0^2/(8\pi G)$ as the usual critical density.  We
thus have a total of $N+2$ parameters: the Hubble constant
$h=H_0/(100~{\rm km/s/Mpc})$, the matter energy density parameter
$\Omega_m$, and the $N$ parameters $f_i$ ($i=1,2,\ldots,N$) describing
the test dark energy function.  The complete set of parameters, then,
is
\begin{equation}
{\bf{s}} \equiv (h, \Omega_m,\,\,\, \rho_i), 
\end{equation}
where $i = 1,...,N$ as described above.
We will vary the number of bins $N$ between 1 and 14,
and look for the optimal fit to the data. To illustrate,
an arbitrary function may become a good approximation to
the data for 4 bins whereas it is a miserable fit for 3 bins.

In \cite{card03} we expanded the adaptive iteration method developed
in Wang \& Garnavich (2001)\cite{Wang01a} and Wang \& Lovelace
(2001)\cite{Wang01b} to include arbitrary time dependence of the dark
energy density; unlike the earlier papers, we do not restrict
ourselves to cases where $\rho_X'(z) \geq 0$.

The adaptive iteration method is designed to optimize the estimation
of the dark energy density $\rho_X(z)$ from data.  It starts with
$f_i=1$ for all $i=1,2,\ldots,N$ (a cosmological constant), and builds
$f_X(z)$ up iteratively while minimizing a modified $\tilde\chi^2$
statistics defined shortly.  This adaptive iteration technique is
further explained in the Appendix.

We can now determine a best fit to the set of parameters $\bf{s}$
by using a $\chi^2$ statistic, with
\cite{SN2}
\be
\chi^2(\mbox{\bf s})=
\sum_l \frac{ \left[ \mu^{(l)}_p(z_l| \mbox{\bf s})-
\mu_0^{(l)}(z_l) \right]^2 }{\sigma_l^2 },
\ee
where $\mu^{(l)}_p(z_l| \mbox{\bf s})$ is the prediction for the
distance modulus at redshift $z_l$, given the set of parameters
$\bf{s}$, and the sum extends over all the observed SNe Ia.  Here
$\sigma_l$ is the dispersion of the measured distance modulus due to
intrinsic and observational uncertainties in SN Ia peak luminosity.

Assuming Gaussian errors, the probability density function for the
parameters ${\bf s}$ is 
\be p(\mbox{\bf s}) \propto \exp\left( -
  \frac{\chi^2}{2} \right).  \ee 
The normalized probability density
function is obtained by dividing the above expression by its sum over
all possible values of the parameters {\bf s}.

The probability density function of a given parameter $s_i$ is
obtained by integrating over all possible values of the other $N+1$
parameters.  To reduce the computation time, we can integrate over the
Hubble constant $h$ analytically, and define a modified $\chi^2$
statistic \cite{Wang01a} as \be
\label{eq:chi2mod}
\tilde{\chi}^2 \equiv \chi_*^2 - \frac{C_1}{C_2} \left( C_1+
  \frac{2}{5}\,\ln 10 \right) - 2 \ln\,h^*.  \ee Here $h^*$ is a
fiducial value of the dimensionless Hubble constant $h$, \ba \chi_*^2
&\equiv& \sum_l \frac{1}{\sigma_l^2} \left( \mu_{p}^{*(l)}-
  \mu_{0}^{(l)} \right)^2, \\
C_1 &\equiv& \sum_l \frac{1}{\sigma_l^2} \left( \mu_{p}^{*(l)}-
  \mu_{0}^{(l)} \right), \\
C_2 &\equiv& \sum_l \frac{1}{\sigma_l^2} , \ea with \be
\mu_p^{*(l)} \equiv \mu_p(z_l; h=h^*)=42.384-5\log h^*+ 5\log
\left[H_0 d_L(z_l)/c\right], \ee The probability distribution function
of the estimated parameters (excluding $h$) is now $\exp\left( -
  \tilde{\chi}^2 /2 \right) $.  It is
straightforward to check that the derivative of $\tilde{\chi}^2$ with
respect to $h^*$ is zero; hence our results are independent of the
choice of $h^*$. We take $h^*=0.65$.

For a given choice of $N$, we can minimize the
modified $\chi^2$ statistic of Eq.(\ref{eq:chi2mod}) to find the best
fit $\Omega_m^{est}$ and $\rho_X(z)$ (parametrized by $\rho_i$, $i=1$, 2, ...,
$N$).  

\subsection{Using the Value of $\Omega_m$ to Constrain $\rho_X'(z)$}

For the remainder of this Section, we restrict ourselves
to Models 1, 2, and 3. We return to Model 4 in the Results Section
below.

To reiterate, we have started from three of the models defined in
Section III: (1) the cosmological constant model with no time
dependence in the energy density, (2) a quintessence model with
$\rho_X'(z)\geq 0$, and (3) an MP Cardassian model with $\rho_X'(z)\leq  0$.
We have constructed a simulated data set for each of these models, and aim
to see how well we can go backwards to determine the sign of
$\rho'(z)$ from this fake data set.  In other words, can we
reconstruct correctly the sign of the time-dependence of the energy
density of the true model?  To do this, we find (via the adaptive
iteration technique) the model that best fits the observed matter
density $\Omega_m$.

For a given number of redshift bins $N$, we can minimize the
modified $\tilde\chi^2$ statistic of Eq.(\ref{eq:chi2mod}) to find the
best fit values $\Omega_m^{\rm est}$ and $f_i^{\rm est}$ for
$\Omega_m$ and $f_X(z)=\rho_X(z)/\rho_X(0)$ parametrized by $f_i$ 
($i=1$, 2, ..., $N$).
For each model in Table 1, we obtain {\it three} sets of best fit
parameters.  We apply three different constraints to the arbitrary
function $\rho_X^{\rm test}(z)$ in order to discover which one allows a
good fit.  The three constraints are: \hfill\break
(i) $\rho_X^{\rm test}(z)=\rho_X^{\rm test}(0)
= {\rm constant}$; i.e.,
a cosmological constant model; \hfill\break
(ii) $d\rho_X^{\rm test}(z)/dz \geq 0$; \hfill\break
and (iii) $d\rho_X^{\rm test}(z)/dz \leq 0$. \hfill\break
We note that the second and third test functions
are monotonically increasing and decreasing respectively,
but do allow portions of the function to be flat as a function of redshift
(i.e., constant for some but not all redshifts).

For each of these three constraints, we find the best fit parameters.
\noindent
If our technique works, the trial that gives the $\Omega_m^{est}$
closest to the true $\Omega_m$ corresponds to the correct underlying
theoretical model (Model 1,2, or 3).  For example, for the case where
Model 2 ($\rho'_X(z)\ge 0$) is the theoretical model, if the trial
with $d\rho_X^{\rm test}(z)/dz \geq 0$ obtains the best value of
$\Omega_m$, then we have reproduced the correct time dependence
of the dark energy density.  Indeed we find that the technique works.

Figure 1 shows our results: panels (a), (b), and (c) correspond to
Models 1, 2, and 3.  For each of the three models, the figure shows
the best fit $\Omega_m^{est}$, under all of the three constraints
above, for $N$ values ranging from 1 to 14.  The different constraints
are represented by different point types.  The dot-dashed horizontal
line is our fiducial value of $\Omega_m = 0.3$ (i.e., we are assuming
that that this is the true value of the matter density), and the solid
horizontal lines indicate 10\% error bars about his fiducial value. We
are assuming that $\Omega_m$ is known to within 10\% from other data
sets.

These plots are NOT intended to emphasize the dependence of
$\Omega_m^{obs}$ on $N$, the total number of redshifts sampled via
linear interpolation (see Eq. (\ref{eq:f(z)})).  Indeed, as discussed
above, the reason that we have found the best fit $\Omega_m$ for a
variety of $N$ values is simply that the parametrization of the
arbitrary function $\rho_X(z)$ may be poor for one value of $N$ but
excellent for another.  For example, two points ($N=1$) are perfectly
adequate to describe a straight line function, but more points are
needed to describe any more complicated function.  Any one value of
$N$ may (by bad luck) give a bad result.  We take a given model to be
a good one if it lies within the 10\% range on $\Omega_m$ for several
values of $N$.
The optimal value of $N$ is the one with the lowest $\tilde\chi^2$,
but we find that $\tilde\chi^2$ does not change much over a wide range
of possible values of $N$.  We look for stability, i.e. for values of
$\Omega_m^{\rm est}$ that do not change much as we vary $N$ slightly.

For Model 2 (with $\rho_X'(z) \geq 0$) and Model 3 (with $\rho_X'(z)
\leq 0$), as $N$ increases, the estimated values $\Omega_m^{\rm est}$
assuming the correct sign of $\rho_X'(z)$ asymptote to the ``true"
value of $\Omega_m=0.3$, while the estimated values $\Omega_m^{\rm
  est}$ assuming the wrong sign of $\rho_X'(z)$ asymptote to an
incorrect value of $\Omega_m$ easily ruled out by observational
constraints on $\Omega_m$.  Indeed our technique works.

For the cosmological constant model (Model 1, top panel of Fig.1), the
estimated $\Omega_m^{\rm est}$ values assuming opposite signs for
$\rho_X'(z)$ fall roughly symmetrically on opposite sides of the
``true" value, and they lie within the error bars on $\Omega_m$.  This
indicates a degeneracy between $\Lambda$ models and dark energy models
that have mildly time-dependent dark energy density.  In other words,
it will be difficult to differentiate a constant dark energy density
from one that has a very small dependence on redshift.  We will
examine this possible degeneracy further in 
Section \ref{sec:6} 
to see how well one can differentiate between these two alternatives.

In conclusion, we can determine the correct sign of $\rho_X'(z)$ if
$\rho_X(z)$ is not too close to a constant. To quantify how close to a
constant we can go, we need to add error bars to the points in figure
1.

\subsection{Using Monte Carlo to Determine Errors}

We evaluate errors by simulating random fluctuations around a fiducial
model. A possible choice of fiducial model would be the input
(theoretical) model we adopted to generate the SN Ia data set in the
first place. However, to be closer to a realistic situation in which
the underlying model has to be determined from data, we will choose
our fiducial model to be the model that best fits the data (for us,
the simulated data).  Specifically, this is the cosmological model
with $\rho_X^{\rm BF}(z)$ (BF = best fit) given by the best-fit values
$\rho_i^{\rm BF}$ ($i=0,1,...,N$) determined above. We have described
in the previous section the method by which we obtain our best fit
dark energy densities for each of Models 1, 2, and 3.  Here, we add
random errors to the ``measured'' distance moduli.

To derive robust error distributions of the estimated parameters
$\Omega_m^{\rm est}$ and $f_i^{\rm est}=\rho_i/\rho_0$ ($i=1$, 2, ..., $N$, see
Eq.(\ref{eq:f(z)})) from each data set, we create $10^3$ Monte Carlo
samples by adding dispersion in peak luminosity of $\Delta
m_{int}=0.16$ mag to the distance modulus $\mu_p(z)$ [see
Eq.(\ref{eq:mu0p})] predicted by the best-fit model (i.e., assuming
that the best-fit model is the true model).  This is equivalent to
making $10^3$ new ``observations'', each similar to the original data
set \cite{Press}.  The same analysis used to obtain the best-fit model
from the data is performed on each Monte Carlo sample.  The
distributions of the resultant estimates of the parameters
($\Omega_m^{\rm est}$ and $f_i^{\rm est}$) can be used to derive the
mean and confidence level intervals of the estimated parameters.  Wang
\& Lovelace \cite{Wang01b} showed that such a Monte Carlo analysis
gives less biased estimates of parameters than a maximum likelihood
analysis, i.e., the Monte Carlo mean of estimated parameters deviate
less from the true values of the parameters.

\section{Results}
\label{sec:6}

\subsection{Models 1, 2, and 3}

Starting from the three dark energy models 1, 2, and 3 in Table 1 with
$\rho_X'(z) =0$, $\rho_X'(z) \geq 0$, and $\rho_X'(z) \leq 0$ respectively,
we obtained simulated data sets. Using these simulated data
sets\footnote{Note that here, once data become available, the
  ``simulated data set'' will be replaced by the real data set. We
  consider three different simulated data sets based on different dark
  energy models, since the nature of dark energy is unknown.}, we
worked backwards to estimate the best fit values of the $N+1$
parameters as defined in the previous section.  In particular, for
each of the simulated data sets we made three trial assumptions, i.e.,
constant, increasing, and decreasing dark energy density.  For each
assumption we found an estimate of $\Omega_m$ and then ran 1000 Monte
Carlo samples to obtain error bars.  The results of this analysis are
shown in Figure 2.  The allowed value of $\Omega_m$, assumed to be
known to within 10\% is shown with arrows: $\Omega_m = 0.3 \pm 0.03$.

Fig.2 confirms the conclusions that we made based on Fig.1.  If the
true dark energy density varies with time (quickly enough) and
monotonically (see Figs.~2b and 2c), the dark energy models with the
wrong sign of $\rho_X'(z)$ or with constant $\rho_X(z)$ are easily
ruled out at $\ga$ 95\% C.L.\ if $\Omega_m$ is known to $\sim$ 10\%
accuracy.  If the true dark energy density is constant with time (Fig.
2a), then while the correct model ($\Lambda$ model) is favored, the
incorrect models (with $\rho_X(z)$ either increasing or decreasing
with redshift) could imitate the correct model if the time variation
in $\rho_X(z)$ is small enough.  We will quantify the size of the
degeneracy region shortly (see Fig.5 and related discussion).

Note that due to computational constraints, we have chosen $N=6$ for
demonstration.  For an actual data set, the Monte Carlo analysis
outlined here should be performed for all values of $N$ (from 1 to a
reasonably large number).

Figure 3 shows the best fit dark energy density for the simulated data
set based on Model 2 (which has $\rho_X'(z)\geq 0$). To obtain figure
3, we take advantage of the fact that our analysis in Fig.2(b) has
allowed us to extract the sign of $\rho_X'(z)$, with the assumption
that we know $\Omega_m$ to 10\% accuracy.  Once the sign of the time
dependence is known, we can deduce the best fit dark energy density
$\rho_X(z)/\rho_X(0)$ shown in Fig. 3, estimated from the Monte Carlo
analysis of the simulated data set based on Model 2.  The solid line
is the true model, i.e., the theoretical curve. One can see that the
dotted points with error bars, obtained from the simulated data using
our technique, match the true model very well. The value of the matter
density corresponding to Fig.3 is $\Omega_m=0.314 (0.271, 0.341)$
(mean and 68.3\% confidence range).  The corresponding figure for
Model 3 ($\rho_X'(z)\leq 0$) was published in \cite{card03}.  Again,
one can see that the time dependence of the energy density can be
determined quite well.

\subsection{Model 4 (linear equation of state)}

Lastly we come to theoretical dark energy models with a linear
equation of state $w_X(z) = w_0 + w_1 z$. Again, we simulate data
based on this underlying model, and again we ask how well we can
determine the time dependence of the energy density of each model.
This linear form of the equation of state is the most common
parametrization used by the community. As stressed previously, one can
extract far more information from data if we aim to learn about the
properties of the underlying energy density, in this case its time
dependence.  In fact, because of our ignorance of the true nature of
dark energy, it is dangerous to rely on specific parametrizations.
Hence, we perform the analysis described in the paper, in which we
treat the dark energy density $\rho_X(z)$ as a free function in
extracting information from the data (in this case simulated data).
We will do a blind test (non-parametric study), in which we do {\it
not} assume the linear form of the equation of state, to see whether
the time-variation of dark energy density can be ascertained.  This
will establish a point of reference with the work by others.  In other
words, we generate models with a given equation of state but then
reconstruct (the time dependence of) the dark energy density itself.

We studied a grid of models, with $ -1.2 \leq w_0 \leq -0.5$, and
$-1.5 \leq w_1 \leq 0.5$, and a grid spacing of $\Delta w_0=0.1$ (0.2
for $w_0\leq -1$) and $\Delta w_1=0.1$.  This represents a total of
147 models.  For each of these theoretical models, we followed the
same procedure described above. First, we simulated data based on each
model; next we obtained the best fit $\rho_X^{\rm test}(z)$ for each
model with the adaptive iteration method (the best fit function chosen
due to its producing $\Omega_m^{est}$ closest to the correct value);
finally we created $10^3$ Monte Carlo samples (fluctuations about the
the best fit function) to obtain error bars.  We followed this
procedure for each of the 147 models.  Fig.~4 shows the estimated
$\Omega_m^{\rm est}$ values (with 1$\sigma$ standard deviations) for a
subset of the $w_X(z)=w_0+w_1 z$ models that we have studied,
assuming: (1) $\rho_X^{\rm test}(z)= {\rm constant}$; (2)
d$\rho_X^{\rm test}(z)/dz\geq 0$ and (3) d$\rho_X^{\rm test}(z)/dz\leq
0$.  The dotted lines in each figure denote $\Omega_m=(0.291,0.309)$,
i.e., assuming that $\Omega_m$ is known to 3\% from other
observations.

Fig. 4 runs through a subset of the grid of theoretical models that we
have studied, namely the models with (a) $w_0 = -1.2$, (b) $w_0 =
-1$, and (c) $w_0 = -0.8$.  In each plot we have selected one 
value of $w_0$ and show
results for a variety of values of $w_1$.  Again, these values of
$w_0$ and $w_1$ correspond to the underlying theoretical model.  Based
on each of these sets of $w_0, w_1$ we simulate data and show how well
we can ascertain the time dependence of the dark energy density.  We
plot the value of $\Omega_m^{est}$ with error bars that results from
our different trial assumptions, as a function of different
(theoretical) values of $w_1$ in the underlying models.  We have
obtained three different $\Omega_m^{est}$'s (for the three different
trial assumptions) for each underlying $w_1$.

In Fig. 4a, we have taken the underlying set of theoretical models to
have $w_X(z) = -1.2 + w_1 z$.  In other words we took $w_0 = -1.2$ and
allowed $w_1$ to vary from $-1.5$ to 0.5 with spacing $\Delta
w_1=0.1$.  Our goal is to obtain the correct value of $\Omega_m^{est}$
only for the test function that has the correct sign of $\rho'_X(z)$.
We see that our technique has indeed reproduced the correct answer,
$\rho_X'(z)\leq 0$ for $w_1 < 0$; however, the answer is ambiguous for
$w_1 >0$, where dark energies that are constant or increasing with
redshift seem to give equally good fits.  In fact we understand the
reason for the success of the technique for some values of $w_1$ and
the failure in other regimes, and explain it here.  As a reminder, we
showed below Eq.(\ref{eq:rhoprime}) that $\rho_X'(z)$ has the same
sign as $1 + w_0 + w_1(z)$.  Hence, for $w_0 = -1.2$, the sign of
$\rho_X'(z)$ is equal to $-0.2 + w_1 z$.  As we discussed in the last
paragraph of Section III, for $w_0<0$ (which is the case for Fig.
4a), the dark energy density is monotonic with redshift for $w_1<0$,
but nonmonotonic for $w_1>0$.  However, as our trial functions we have
only considered monotonic dark energy densities.  Hence it is not
surprising that our technique only reproduces the correct time
dependence of the dark energy in the regimes of parameter space that
correspond to dark energies that are monotonically increasing or
decreasing in time.  The technique works within the regime of validity
of our trial assumptions.

In Fig.  4b, the underlying model is $w_X(z) = -1 + w_1 z$, again for
$w_1$ ranging from $-1.5$ to 0.5. In this case the corresponding dark
energy density is monotonic with redshift for all values of $w_1$, and
the technique reproduces the right answer for all $w_1$. 

In Fig. 4c, the underlying model is $w_X(z) = -0.8 + w_1 z$, again for
$w_1$ ranging from $-1.5$ to 0.5. 
This corresponds to a dark energy
density that is monotonic for $w_1>0$ but non-monotonic for $w_1<0$.
The dark energy density grows with redshift until $z_{crit} = (1+
w_0)/|w_1| = 0.2/|w_1|$ (as obtained from Eq.(\ref{eq:zcrit})) and
then decreases with redshift.  However, the value of $z_{crit}$ is
so small in most models as to be irrelevant; i.e. the function is
essentially monotonically decreasing over most of the redshifts at
which supernova data are taken.  For example, for $w_1 = -1.5$,
$z_{crit} = 0.133$.  As $w_1$ approaches 0, the value of
$z_{crit}$ grows; clearly at $w_1 = -0.2$ and $z_{crit}=1$, the
turnover from a growing dark energy density to a decreasing one takes place in
the middle redshift of the supernova data.  In that case we would
expect a constant trial with $\rho_X'(z) = 0$ to be no worse a fit than a
monotonically increasing or decreasing one.  In other words, a
cosmological constant appears to fit the data, although of course it
is the wrong answer.  Indeed this is borne out by the results of Fig.
4c.  Here, our technique is very successful at reproducing the correct
sign of the time dependence of the dark energy in those regimes where
it is sensible to approximate the dark energy density as being
monotonic in time.

Models with $w_0 = -0.9, -0.7, -0.6, \,\, {\rm and} \,\,
-0.5$ behave similarly to those shown in Fig.4.  As $w_0$
gets less negative, the value of $z_{crit}$ (at which the derivative
of the dark energy density changes from positive to negative) gets
bigger.
Again, our technique works to obtain the correct time dependence of
the dark energy density if it is monotonic in time.

To quantify how much the $\rho_X'(z) \neq 0$ models differ from the
$\Lambda$ models, we have defined a quantity $Q$ as the number of
standard deviations in the difference of the average estimated value
of $\Omega_m$ with constant and non-constant dark energy density,
\begin{equation}
\label{eq:Q}
Q \equiv \frac{ \Biggl| \Bigl\langle \Omega_m^{\rm est}
  \Big\vert_{\rho_X={\rm non-const}} \Bigr\rangle - \Bigl\langle
  \Omega_m^{\rm est}\Big\vert_{\rho_X={\rm const}} \Bigr\rangle
  \Biggr|} {\sqrt{ \Bigl(\Delta \Omega_m^{est}\Big\vert_{\rho_X={\rm
        non-const}} \Bigr)^2 +\Bigl(\Delta
    \Omega_m^{est}\Big\vert_{\rho_X={\rm const}} \Bigr)^2 } } .
\end{equation}
The thick solid line in the bottom panel of each section of
Fig.4 shows $Q$ as a function of $w_1$ for various $w_0$ values.  As
there are two test functions with time varying dark energy density,
$\rho_X'(z) \geq 0$ and $\rho_X'(z) \leq 0$, in the figures we plot the
quantity $Q$ for that time varying function that produces the value of
$\Omega_m^{est}$ that best fits the data.

We note that our work is quite different to prior studies performed by
other authors.  Previous work (e.g.,
\cite{mbs,barger,hutturn,Linder}) has attempted to examine how
well one can reconstruct the dark energy equation of state,
particularly if one assumes it has the form $w_X(z)=w_0+w_1 z$.  Our
study differs from previous studies in being {\it non-parametric}
(hence not dependent on $w_X(z)=w_0+w_1 z$ being the correct model,
but only using it as a test case), and spanning a continuous parameter
space in $(w_0, w_1)$ (via smooth interpolations in our grid of 147
models).

\vskip 0.2in
\centerline{\bf How Small Can We Make the Region of Degeneracy?}
\vskip 0.1in

We can ask the question: what is the range of ($w_0$, $w_1$) over
which one cannot determine the sign of the time dependence?  In other
words, what is the range of degeneracy between a cosmological constant
and a time-changing vacuum energy appearing to fit the data equally
well?  And how far can it be shrunk down?

Fig.5 shows the ($w_0$, $w_1$) parameter space that we have studied.
The models that lie within the shaded region cannot be differentiated
from a $\Lambda$ model even if $\Omega_m$ is known to 1\%.  Similarly,
models which lie within the dotted, dashed, and solid lines cannot be
differentiated from a cosmological constant model if $\Omega_m$ is
known to within 3\%, 5\%, and 10\% accuracies respectively.

The degeneracy region shown in Fig.5 is centered about the line
\be
\label{eq:central}
1+w_0 \simeq - \frac{z_{max}}{3}\, w_1 , 
\ee
where $z_{max}$ is the
maximum redshift of the survey ($z_{max}=1.7$ for SNAP).

In the parameter space outside of the shaded region in Fig.5,
$\rho_X'(z) \neq 0$ models are preferred over the $\Lambda$ models if
$\Omega_m$ is known to 1\% accuracy, indicating our ability to detect
the time-variation of the dark energy density at 1$\sigma$ or higher
significance levels.

Finally, we point out a way to reduce the degeneracy by taking
advantage of the fact that the slope in Eq.(\ref{eq:central}) changes
for different values of $z_{max}$.  Hence changing $z_{max}$ will
rotate the degeneracy region in $(w_0, w_1)$.  One can choose a
variety of different values of $z_{max}$ to break the degeneracy. In
other words, in addition to using the entire data set, one can
restrict the data out to a variety of different cutoff redshifts to
obtain complementary information.  If, in addition to the full data
set, we consider only those data to a second maximum redshift
$z_{2,max}$, we can reduce the degeneracy region in Fig.5
significantly. In Figure 5 we have plotted both the degeneracy region
obtained using all the data, as well as dot-dashed lines illustrating
the different degeneracy region if only data out to $z_{2,max}=0.5$
were used.\footnote{These are qualitative illustrations, not from
  actual calculations (which would involve lengthy computations).}
The combination of information from these dot-dashed lines together
with the shaded region allow us to break the degeneracy substantially.
The shaded region in Fig.5 bounded by dot-dashed lines illustrates the
smallest possible degeneracy region if $\Omega_m$ is known to 1\%.

\section{Conclusions}
\label{sec:7}

We have investigated how well future supernova data from dedicated
telescopes (such as SNAP), when combined with data of nearby
supernovae, can be used to determine the time dependence of the dark
energy density. For quantitative comparison, we have done an extensive
study of a number of dark energy models, with dark energy density that
is constant, increasing, and decreasing in time. Based on these models
we have simulated data sets in order to show that we can indeed
reconstruct the correct sign of the time dependence of the dark energy
density.

Among the dark energy models we studied are those parametrized by an
equation of state $w_X(z)=w_0+w_1 z$.  Here, $w \equiv p/\rho$.  We
studied a grid of 147 models, for $ -1.2 \leq w_0 \leq -0.5$, and
$-1.5 \leq w_1 \leq 0.5$.  We emphasize that it is the dark energy
density that we reconstructed, {\it not} the equation of state.  We
find that there is a degeneracy region in the ($w_0, w_1$) parameter
space centered near $1+w_0 = -w_1 z_{max}/3$ (where $z_{max}$ is the
maximum redshift of the survey, e.g., $z_{max}=1.7$ for SNAP); the
models that lie within this region cannot be differentiated from a
$\Lambda$ model even if $\Omega_m$ is known independently to 1\%
accuracy (we compute the size of the region for $\Omega_m$ known to
varying degrees of accuracy).  Outside of this degeneracy region, we
can detect the time variation of the dark energy density at 1$\sigma$
or higher significance levels.

We emphasize that, given the same data, one can learn much more by
reconstructing the dark energy density directly (and its time
dependence) than by attempting to reconstruct its equation of state.

\acknowledgments It is a pleasure for us to thank Greg Tarle for
helpful comments.  We acknowledge support from 
NSF CAREER grant AST-0094335 (Y.W. and V.K.); 
the Department of Energy grant at the
University of Michigan and the Michigan Center for Theoretical Physics
(K.F.);
the DOE at Fermilab
and Chicago, from NASA grant NAG5-10842 at
Fermilab, and from the NSF Center for Cosmological
Physics at Chicago (J.A.F).

\bigskip\bigskip

{\bf APPENDIX: ADAPTIVE ITERATION TECHNIQUE}

\bigskip

The goal of the adaptive iteration technique is to reconstruct (as
accurately as possible) the function $f_X(z) = \rho_X(z)/\rho_X(0)$
from a simulated data set.  We start from a time independent function
(in which $f_X(z) = 1$ for all z) and build up the function
iteratively to find that function which best matches the data.

To illustrate the adaptive iteration technique, we present an example.
In this appendix we here restrict our discussion to monotonically
increasing forms of $f_X(z)$.  Let us consider the case where we break
up the function (and the simulated data) into five equally spaced bins
in redshift space, i.e., $N=5$ so that $z$ ranges from
$0,z_1,z_2,z_3,z_4,z_5$.  We will start with a flat function,
$f_X(z)=\rho_X(z)/\rho_X(0) = 1$ and build up from there. In all the
iterations, we will always keep $f_X(0)=\rho_X(z=0)/\rho_X(0) = 1$
fixed (it's an identity equation since $\rho_X(0)\equiv \rho_X(z=0)$),
and vary $f_X(z)$ at the other
values of $z$. Here is how we proceed in the first iteration:\\
\noindent
(1) Compute $\chi^2$ for $f_X(z)=1$.\\
(2) Increase $f_X(z_i)$ for all $i=1,2,3,4,5$ by one stepsize,
$\Delta$, i.e., $f_X(z_i)=1+\Delta$ for all $i=1,2,3,4,5$.  Compute
$\chi^2$. If the current $\chi^2$ is smaller than the previous
$\chi^2$ (from the previous step), the new $f_X(z_i)$ values are
favored; keep them.  Otherwise, the previous values of $f_X(z_i)=1$
are favored.  As stepsize we used primarily $\Delta=0.01$, but also
$\Delta = 0.05$ and $0.1$ for comparison. We find that the results
are not sensitive to the stepsize; of course,
the smaller the stepsize, the longer the running time.\\
(3) Increase $f_X(z_i)$, $i=2,3,4,5$ by one stepsize, $\Delta$, i.e.,
$f_X(z_i)=f_X(z_i)^{prev}+\Delta$, $i=2,3,4,5$, where
$f_X(z_i)^{prev}$ are the $f_X(z_i)$ values favored by the previous
step.  Compute $\chi^2$. If the current $\chi^2$ is smaller than the
previous $\chi^2_{min}$ (from the previous step), the new $f_X(z_i)$
values are favored; keep them.
Otherwise, $f_X(z_i)^{prev}$ are favored.\\
(4) Increase $f_X(z_i)$, $i=3,4,5$ by one stepsize, $\Delta$, i.e.,
$f_X(z_i)=f_X(z_i)^{prev}+\Delta$, $i=3,4,5$, where $f_X(z_i)^{prev}$
are the $f_X(z_i)$ values favored by the previous step.  Compute
$\chi^2$. If the current $\chi^2$ is smaller than the previous
$\chi^2_{min}$ (from the previous step), the new $f_X(z_i)$ values are
favored; keep them.
Otherwise, $f_X(z_i)^{prev}$ are favored.\\
(5) Increase $f_X(z_i)$, $i=4,5$ by one stepsize, $\Delta$, i.e.,
$f_X(z_i)=f_X(z_i)^{prev}+\Delta$, $i=4,5$, where $f_X(z_i)^{prev}$
are the $f_X(z_i)$ values favored by the previous step.  Compute
$\chi^2$. If the current $\chi^2$ is smaller than the previous
$\chi^2_{min}$ (from the previous step), the new $f_X(z_i)$ values are
favored; keep them.
Otherwise, $f_X(z_i)^{prev}$ are favored.\\
(6) Increase $f_X(z_i)$, $i=5$ by one stepsize, $\Delta$, i.e.,
$f_X(z_i)=f_X(z_i)^{prev}+\Delta$, $i=5$, where $f_X(z_i)^{prev}$ are
the $f_X(z_i)$ values favored by the previous step.  Compute $\chi^2$.
If the current $\chi^2$ is smaller than the previous $\chi^2_{min}$
(from the previous step), the new $f_X(z_i)$ values are favored; keep
them.  Otherwise, $f_X(z_i)^{prev}$ are favored.\\

In the second iteration, repeat steps (2)-(6), but with these changes
in step 2: replace $f_X(z_i)=1+\Delta$ with
$f_X(z_i)=f_X(z_i)^{prev}+\Delta$, replace ``the previous $\chi^2$"
with ``the previous $\chi^2_{min}$", and replace ``the previous values
of $f_X(z_i)=1$" with ``$f_X(z_i)^{prev}$".

Subsequent interations follow the same procedure as the second
iteration. Continue the interations until $\chi^2_{min}$ stops
changing.

We successively perform further iterations to ascertain the function
$\rho_X(z)$ that best fits the data. As described here, one can only
end up with a monotonically increasing form for $\rho(z)$.  For
monotically decreasing function, the procedure is exactly the same,
with $+\Delta$ replaced by $-\Delta$.


\clearpage

\setcounter{figure}{0}
\begin{figure}
\psfig{width=\textwidth,file=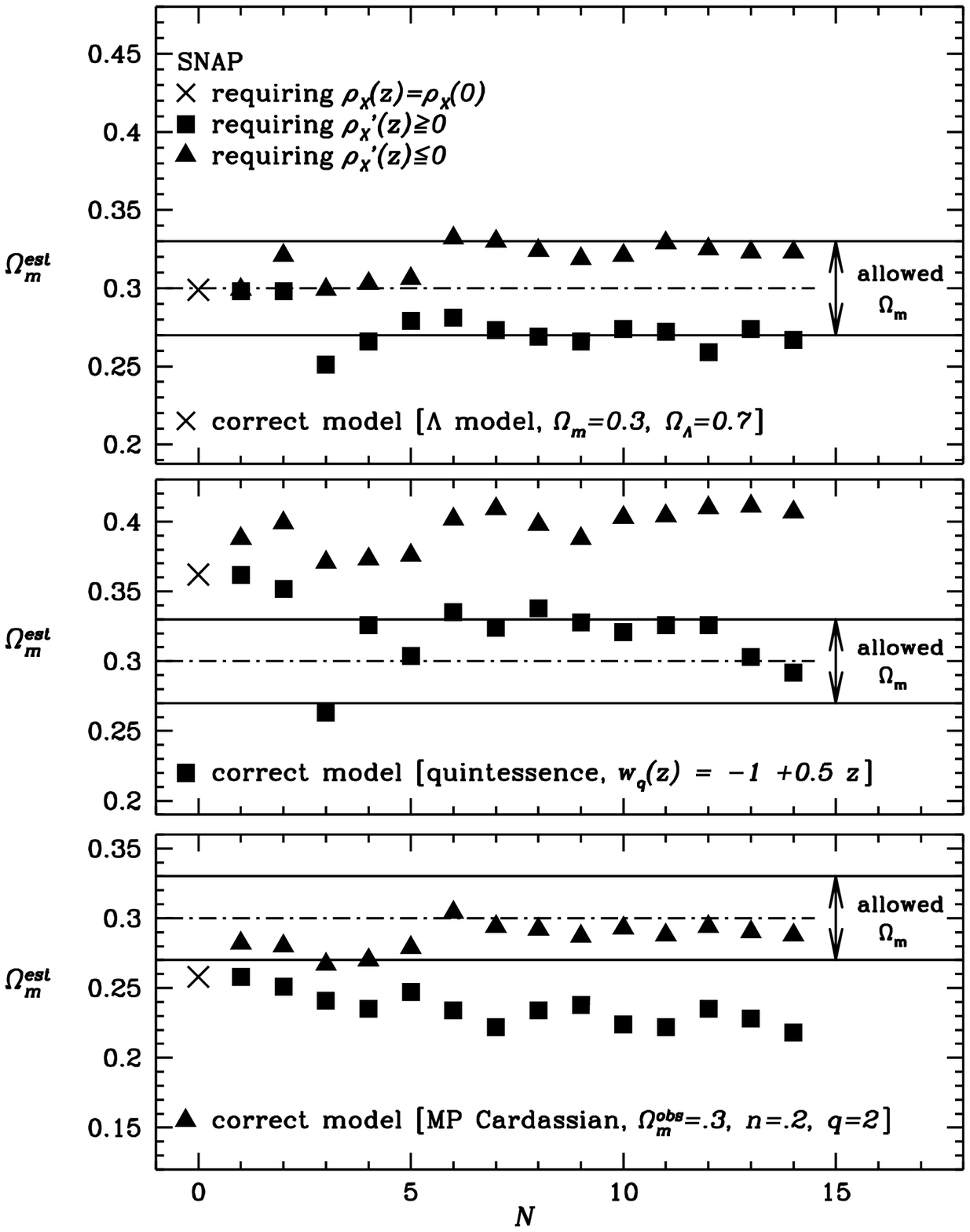} 
\caption[f1.eps]
{The estimated best fit $\Omega_m^{est}$, shown for a variety of
  values of the number of parameters $N$. The three panels are for the
  dark energy models 1, 2, and 3 (given in Table 1) respectively.
  Each panel plots the estimated $\Omega_m^{est}$ for three trial
  functions: crosses indicate constant $\rho_X^{test}(z)$, squares
  indicate $d\rho_X^{test}/dz > 0$, and triangles indicate
  $d\rho_X^{test}/dz < 0$.  The dot-dashed line indicates the ``true"
  value of $\Omega_m=0.3$ in each model, the solid lines indicate the
  $\pm$10\% range of $\Omega_m$.  In each panel the test function with
  the correct time dependence (same as the underlying theoretical
  model) produces an acceptable $\Omega_m^{est}$ that matches data;
  the test function with the wrong time dependence produces an
  incorrect value of $\Omega_m$ and hence can be ruled out.  This
  technique can reproduce the correct time dependence of the dark
  energy density. }
\end{figure}

\newpage

\setcounter{figure}{1}
\begin{figure}
\psfig{width=\textwidth,file=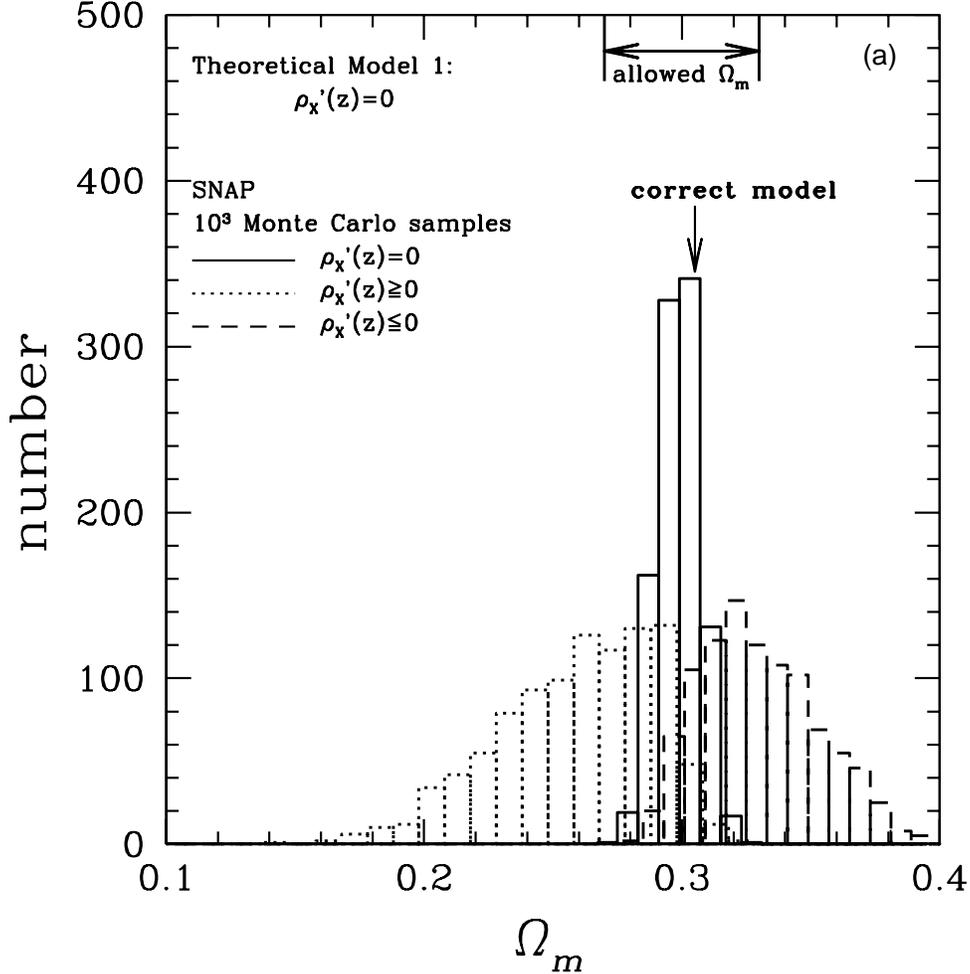} 
\caption[f2a.eps]
{(a) Distributions of the estimated $\Omega_m$ values from $10^3$
  Monte Carlo samples.  Here, the underlying theoretical model is a
  cosmological constant.  The Monte Carlo samples are obtained under
  three different trial assumptions about the time dependence of the
  dark energy $\rho_X'(z)$: increasing, decreasing, or constant.  The
  allowed value of $\Omega_m = 0.3 \pm 0.03$ is indicated.  The
  correct value of $\Omega_m$ is reproduced for the correct time
  dependence of the dark energy, $\rho_X'=0$; hence one is led to
  conclude that the underlying model is a cosmological constant.
  Note, however, that for this case some of the models with time
  dependent $\rho_X$ also produce the right values of $\Omega_m$, so
  that there is some degeneracy.}
\end{figure}

\newpage

\setcounter{figure}{1}
\begin{figure}
\psfig{width=\textwidth,file=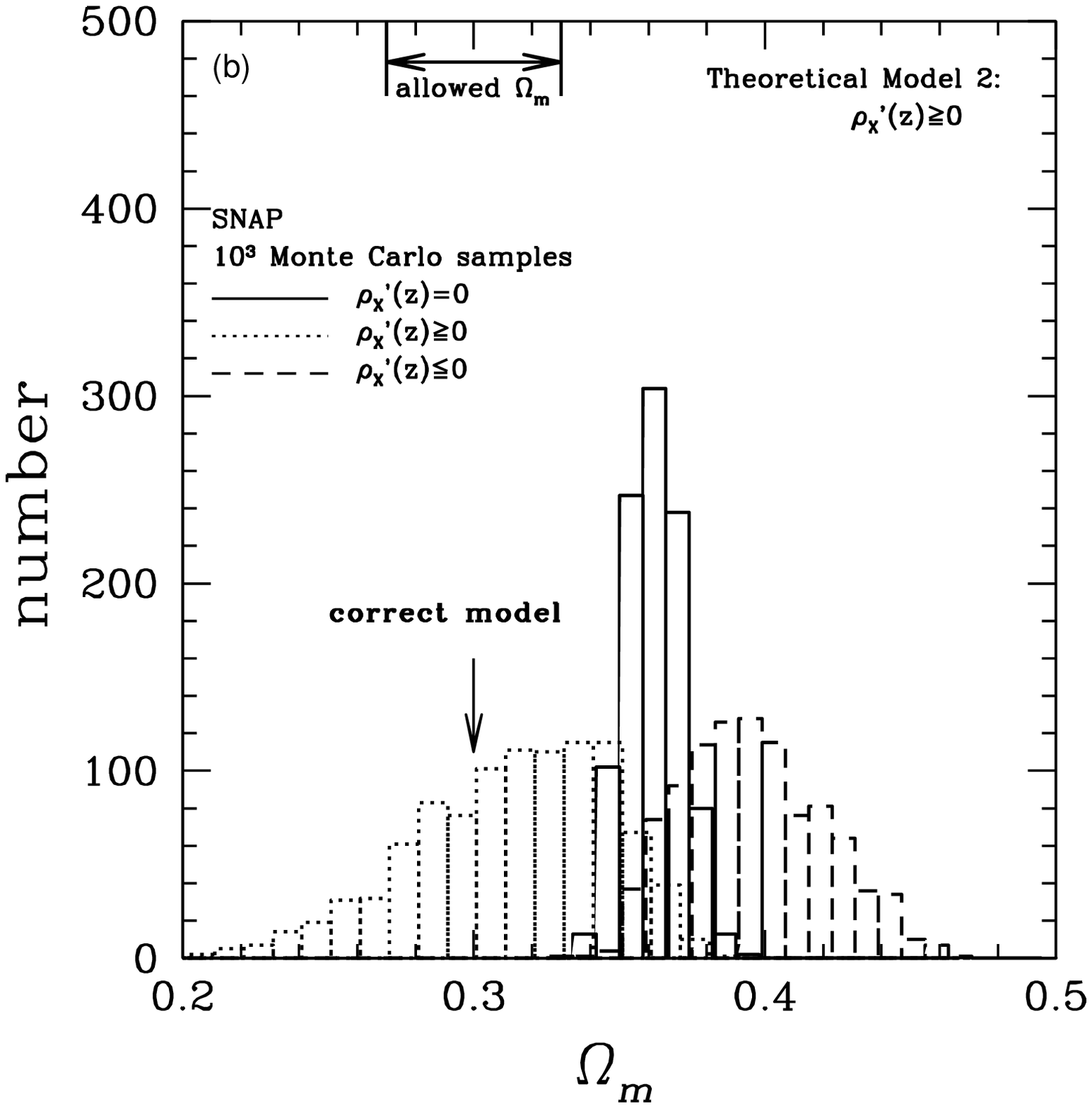} 
\caption[f2b.eps]
{b) Same as Fig.2(a), but for the simulated data set based on Model 2.
  Upon running $10^3$ Monte Carlo samples, the correct value of
  $\Omega_m$ is reproduced only for the trial functions with
  $\rho_X'>0$, which agrees with the time dependence of the underlying
  model. [The underlying model is a quintessence model with $w_x(z) =
  -1 + 0.5 z$ and $\rho_X'(z) > 0$.] We have taken $N=6$ for
  illustration. Hence one recovers the correct time dependence of the
  energy density.}
\end{figure}

\newpage

\setcounter{figure}{1}
\begin{figure}
\psfig{width=\textwidth,file=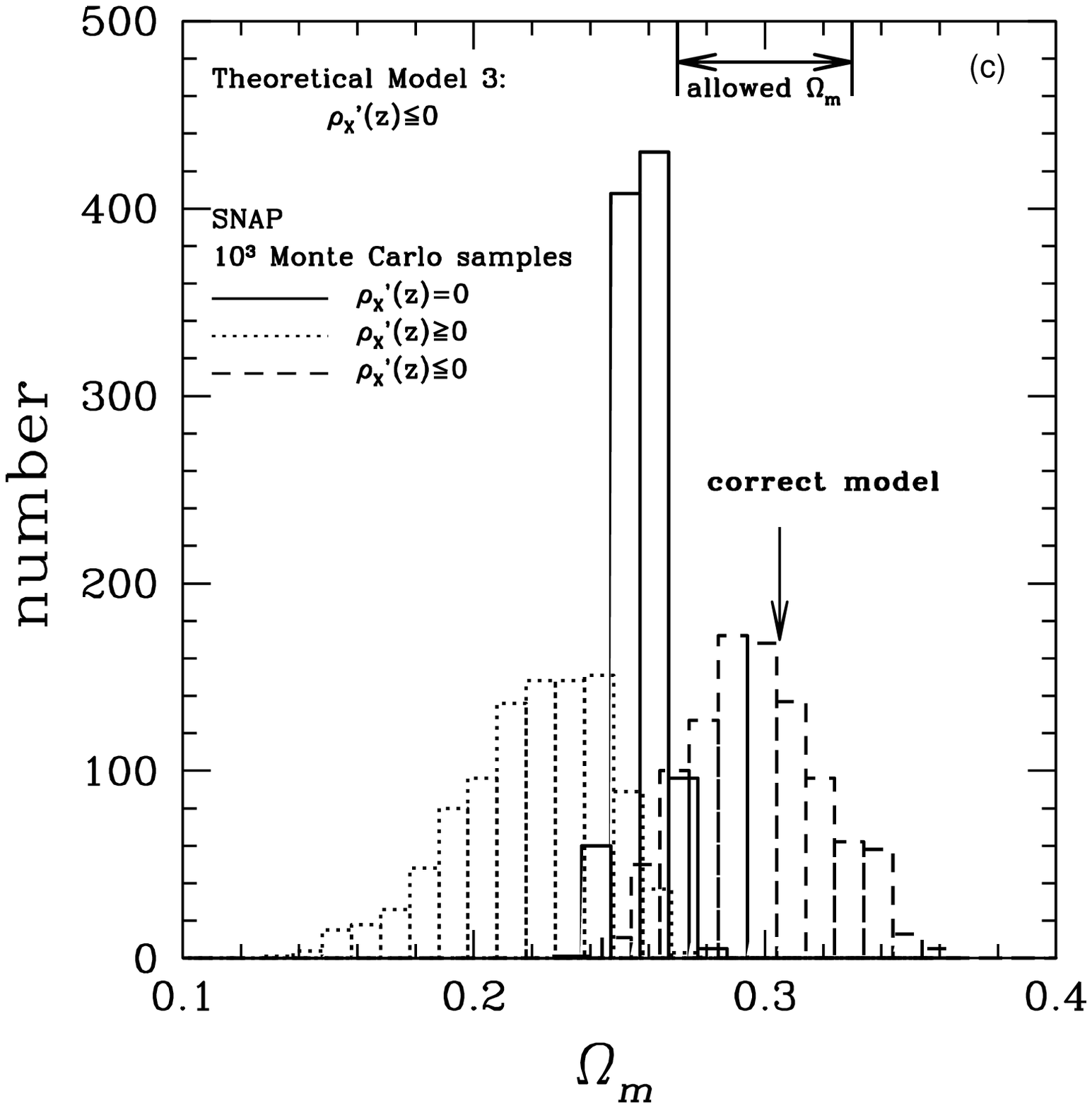} 
\caption[f2c.eps]
{c) Same as Fig.2(a), but for the simulated data set based on Model 3.
  Upon running $10^3$ Monte Carlo samples, the correct value of
  $\Omega_m$ is reproduced only for the trial functions with
  $\rho_X'<0$, which agrees with the time dependence of the underlying
  model. [The underlying model is a MP Cardassian model with n=0.2 and
  q=2, and $\rho_X'(z) <0$.] We have taken $N = 6$.  Hence one again
  recovers the correct time dependence of the energy density.}
\end{figure}

\newpage

\setcounter{figure}{2}
\begin{figure}
\psfig{width=\textwidth,file=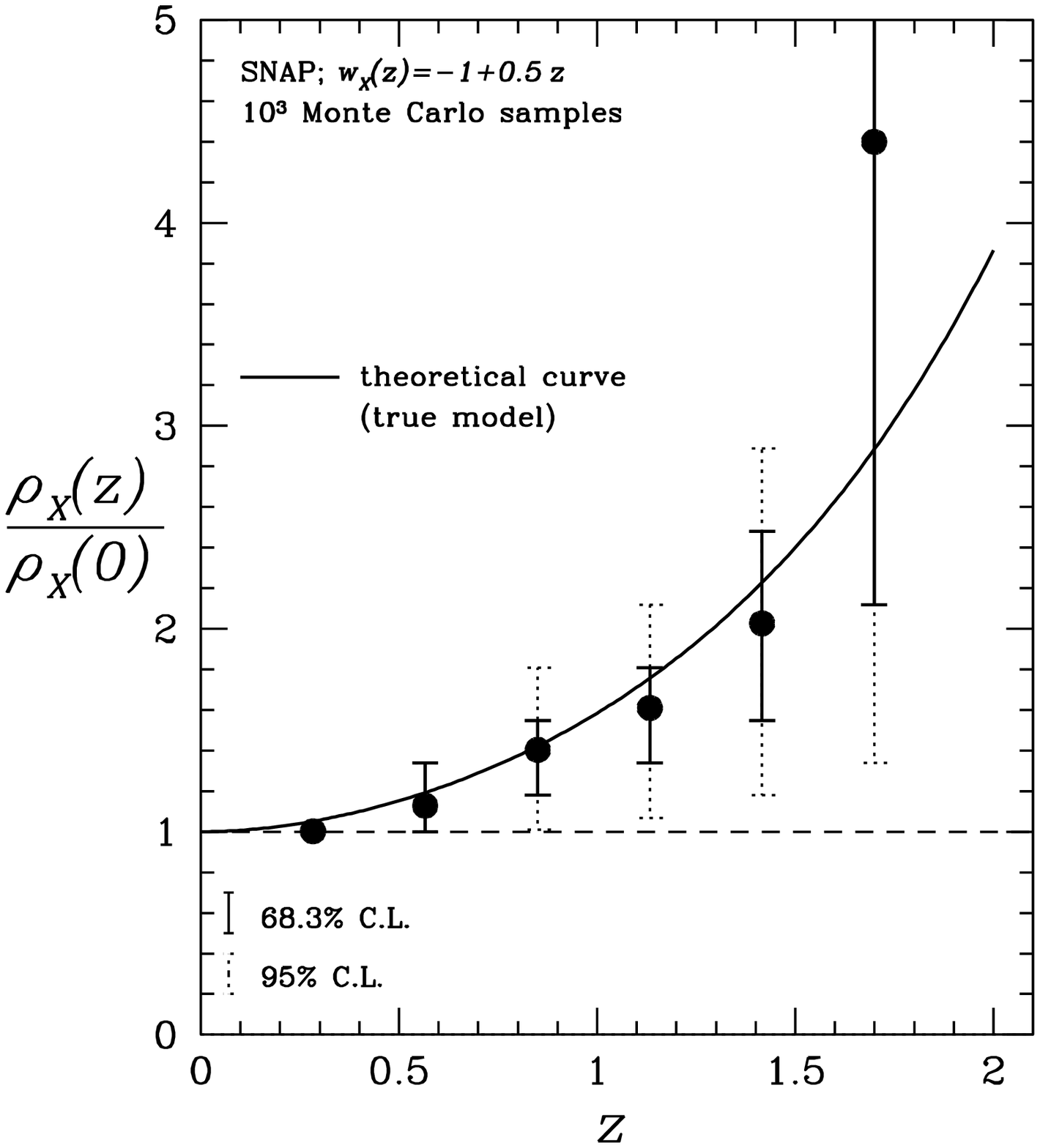} 
\caption[f3.eps] {Best fit dark energy density $\rho_X(z)/\rho_X(0)$
estimated from the Monte Carlo analysis of the simulated data set
based on Model 2, which has $\rho_X'(z)>0$. This plot assumes that
$\Omega_m$ is known to 10\% accuracy [and that the sign of the time
dependence has been extracted as shown in Figure 2]. The solid line is
the true model, i.e., the theoretical curve. One can see that the
dotted points with error bars, obtained from the simulated data using
our technique, match the true model very well.  Here, we have taken
$N=6$.  }
\end{figure}

\newpage

\setcounter{figure}{3}
\begin{figure}
\psfig{width=\textwidth,file=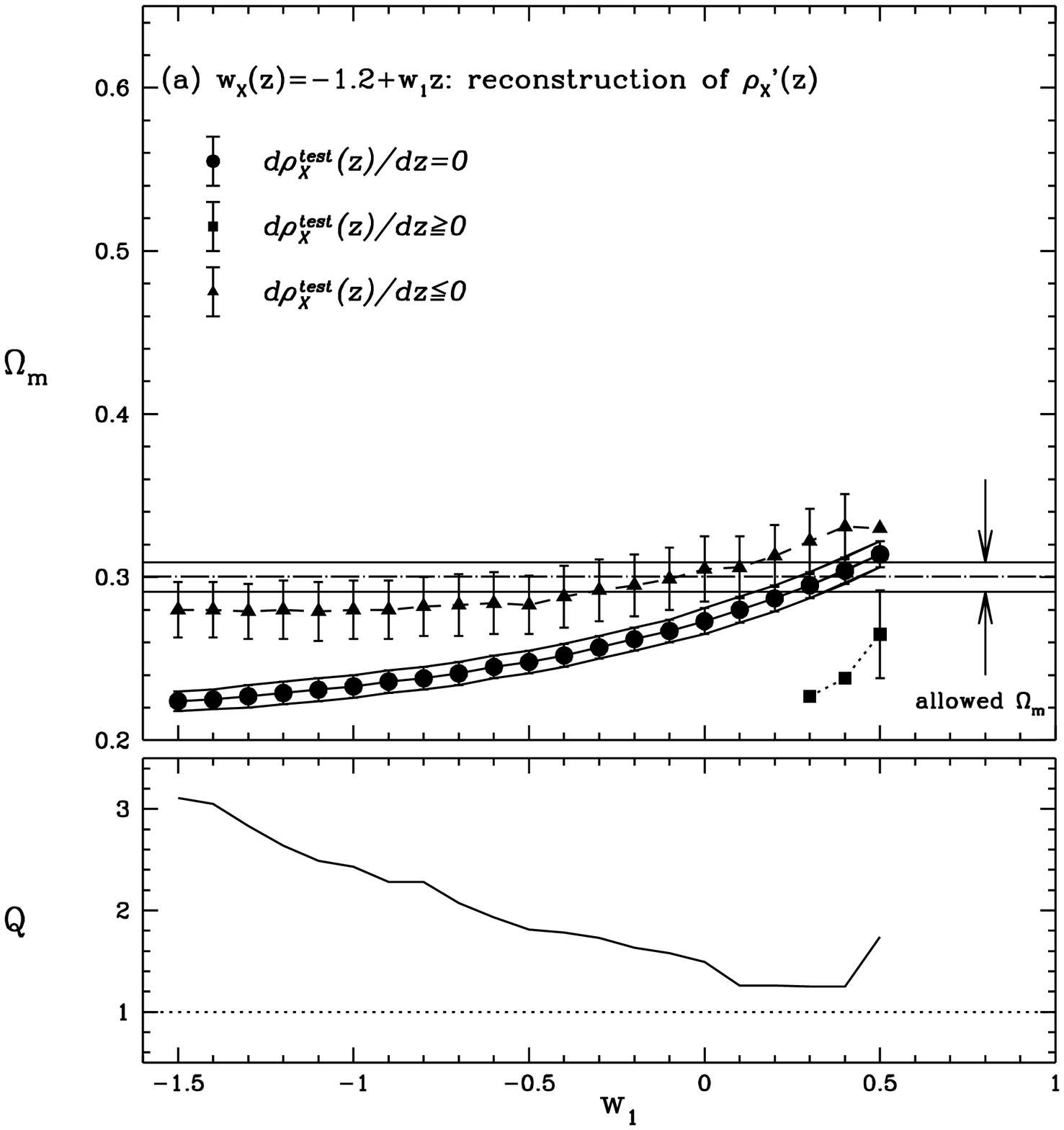} 
\caption[f4a.eps]
{ The estimated $\Omega_m$ values (with 1$\sigma$ standard deviations)
  for $w_X(z)=w_0+w_1 z$ models.  In each plot in panels a-g, we have
  selected one value of $w_0$ and show results for a variety of values
  of $w_1$; these values of $w_0$ and $w_1$ correspond to the
  underlying theoretical model.  Based on each of these sets of $w_0,
  w_1$ we simulated data, and in the upper half of each panel,
plotted the values of $\Omega_m^{est}$
  with error bars that result from our three different trial
  assumptions, as a function of different (theoretical) values of
  $w_1$ in the underlying models.  The three different trial
  assumptions are represented by: circles for $\rho_X^{\rm test} =
  {\rm constant}$, squares for $d\rho_X^{\rm test}/dz \geq 0$, and
  triangles for $d\rho_X^{\rm test}/dz \leq 0$.

The dotted lines denote $\Omega_m=(0.291,0.309)$ (here, we assume that
$\Omega_m$ is known to 3\% from other observations.)  In the lower
half of each panel, we have plotted the quantity $Q$ defined in
Eq.(\ref{eq:Q}), the number of standard deviations in the difference
of the average estimated value of $\Omega_m$ with constant and
non-constant dark energy density. 

\hfill\break
(a) Underlying theory: $w_0 = -1.2$ and
  $-1.5 \leq w_1 \leq 0.5$. The correct value of $\Omega_m$ is
  obtained for $\rho_X'(z) \leq 0$ for $w_1 < 0$; here our technique has
  indeed reproduced the correct time dependence of the dark energy.
  However, the answer is ambiguous for $w_1 >0$, where the underlying
  theory is non-monotonic.}
\end{figure}

\newpage

\setcounter{figure}{3}
\begin{figure}
\psfig{width=\textwidth,file=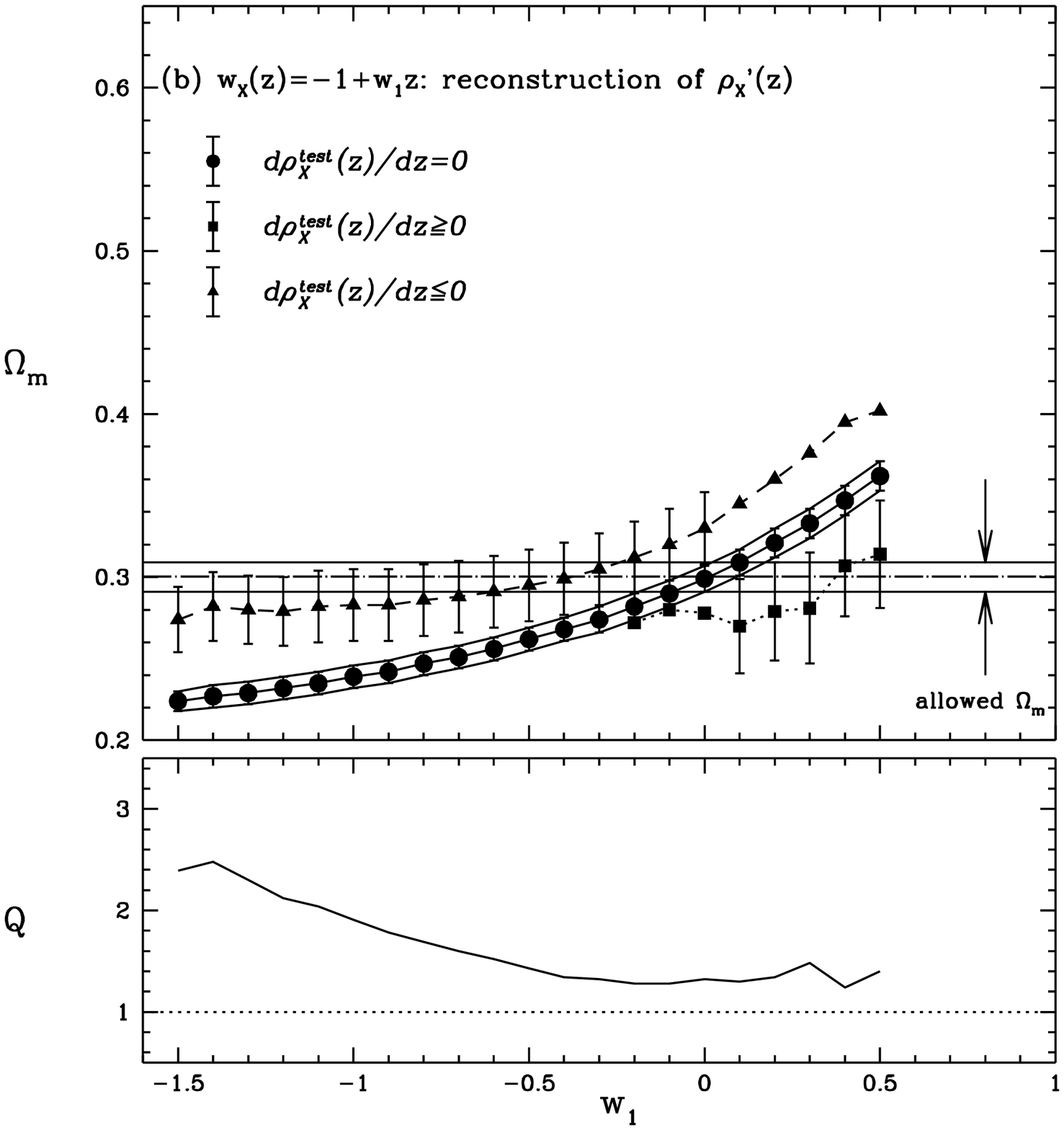} 
\caption[f4b.eps]
{(b) Underlying theory: $w_0=-1$ and $-1.5\leq w_1 \leq 0.5$. 
  Here our technique obtains an allowed value of $\Omega_m$ and hence
  reproduces the correct time dependence of the dark energy for all
  $w_1$.
}
\end{figure}

\newpage

\setcounter{figure}{3}
\begin{figure}
\psfig{width=\textwidth,file=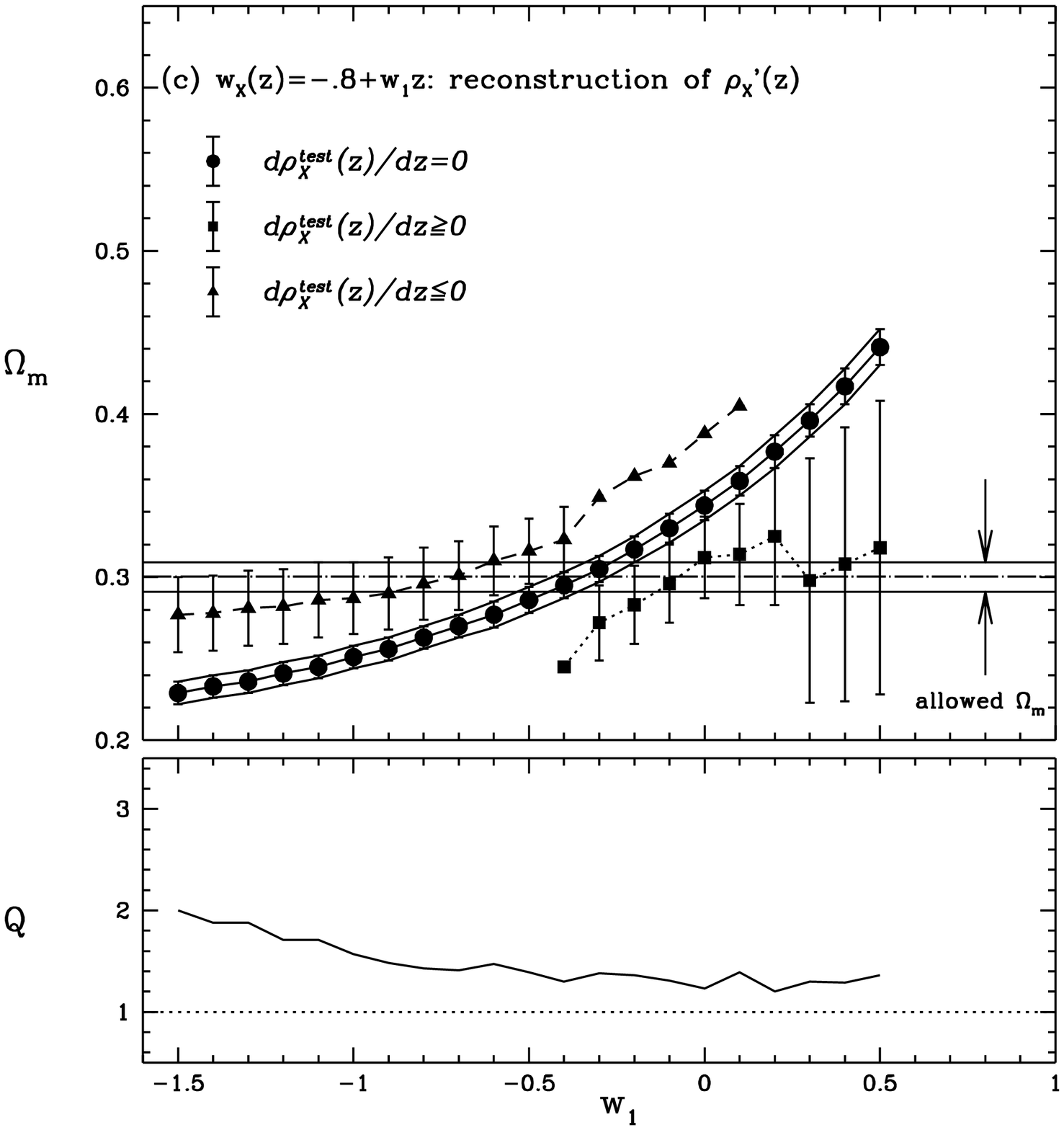} 
\caption[f4d.eps]
{(c) Underlying theoretical model: $w_0=-0.8$ and $-1.5 \leq w_1 \leq 0.5$
(see discussion in text).
}
\end{figure}

\setcounter{figure}{4}
\begin{figure}
\psfig{width=\textwidth,file=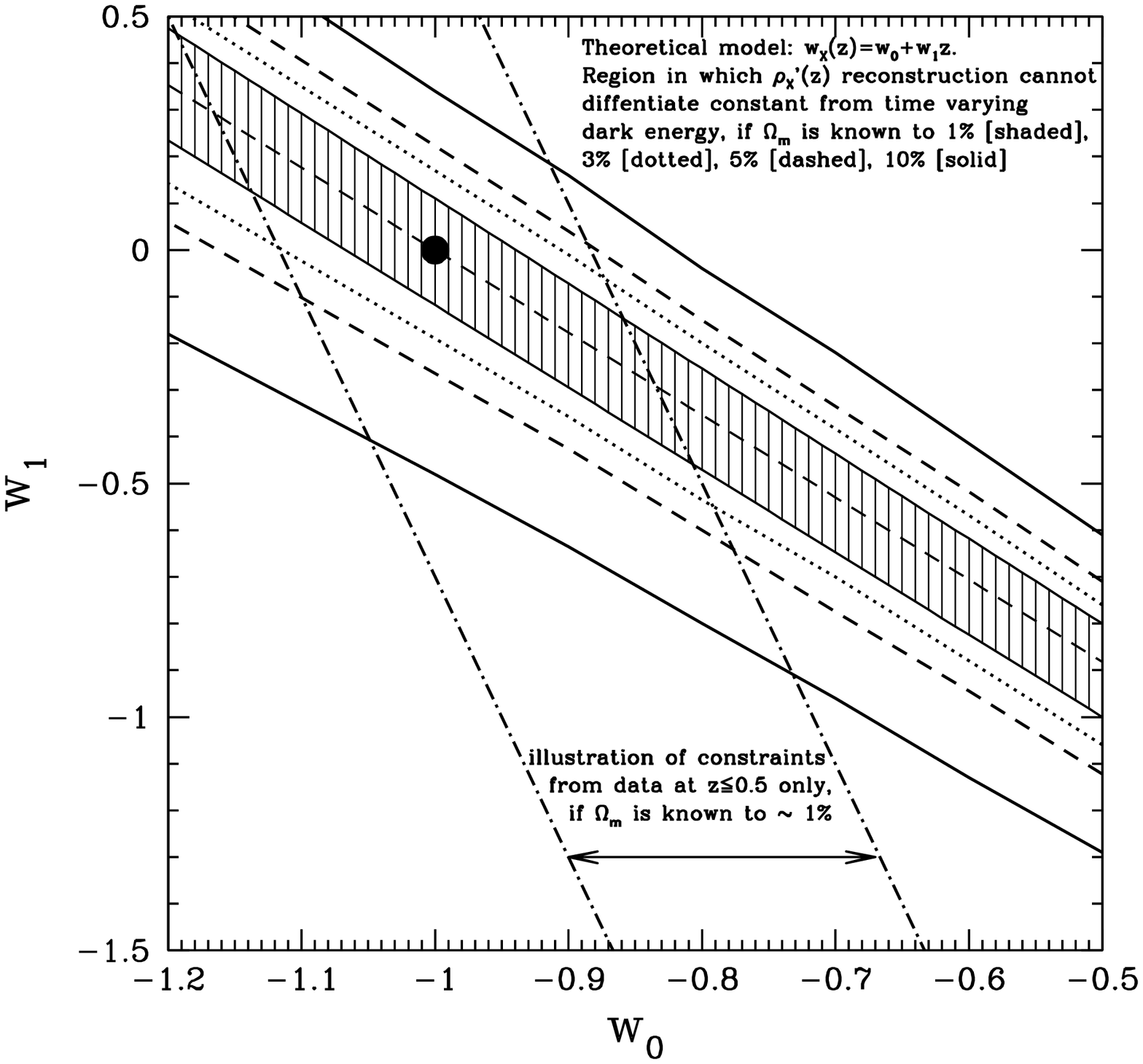} 
\caption[f5.eps]
{The ($w_0$, $w_1$) parameter space that we have studied.  For the
  theoretical models with $w_0$ and $w_1$ that lie within the shaded
  region, the reconstructed $\rho_X'(z)$ cannot be differentiated from
  a $\Lambda$ model (indicated by a fat circle) even if $\Omega_m$ is
  known to 1\%.  Outside of this region, the sign of $\rho_X'(z)$ can
  clearly be ascertained (if $\Omega_m$ is known to 1\%). Similarly,
  models which lie within the dotted, dashed, and solid lines cannot
  be differentiated from a constant $\rho_X'(z) = 0$ model if
  $\Omega_m$ is known to within 3\%, 5\%, and 10\% accuracies
  respectively.  The degeneracy region is centered about the line
  $1+w_0 \simeq - \frac{z_{max}}{3}\, w_1 $, where $z_{max}$ is the
  maximum redshift of the survey ($z_{max}=1.7$ for SNAP). The
  dot-dashed line illustrates (roughly) the different degeneracy
  region if only those data out to a cutoff redshift of 0.5 were used,
  if $\Omega_m$ were known to 1\%. Note that the degeneracy can be
  reduced by examining different portions of the data out to different
  redshifts.  }
\end{figure}


\begin{references}

\bibitem{wmap} C. Bennett, et al. (the WMAP team), astro-ph/0302208;
  D. Spergel, et al. (the WMAP team), Astrophys.J.Suppl. {\bf 148},
  175 (2003).
  
\bibitem{yaya} K. Freese, F.C. Adams, J.A. Frieman, \& E. Mottola,
  Nucl. Phys. {\bf B287}, 797 (1987); P.J.E. Peebles \& B. Ratra, ApJ
  {\bf 325L}, 17 (1988); C. Wetterich, Nucl. Phys. {\bf B302}, 668
  (1988); J. Frieman, C. Hill, A. Stebbins, and I. Waga,
  Phys.Rev.Lett. {\bf 75}, 2077 (1995), I. Zlatev, L. Wang, and P.J.
  Steinhardt, Phys. Rev. Lett.  {\bf 82}, 896 (1999).
  
\bibitem{card} K. Freese, and M. Lewis, Phys.Lett. {\bf B540}, 1
  (2002); K. Freese, Nuclear Physics B (Proc. Suppl.) {\bf 124}, 50
  (2003)
  
\bibitem{mpcard} P. Gondolo and K. Freese, Phys.Rev. {\bf D68} 063509
  (2003)
  
\bibitem{card03} Y. Wang, K. Freese, P. Gondolo, and M. Lewis,
  Astrophys.J. {\bf 594}, 25 (2003).
 
\bibitem{others} L. Parker and A. Raval, Phys. Rev. D {\bf 60}, 063512
  (1999); C. Deffayet, Phys. Lett. {\bf B502}, 199 (2001); N. Bilic,
  G.B Tupper, and R. Viollier, Phys.Lett. {\bf B535} 17 (2002); M.
  Ahmed, S. Dodelson, P.B. Greene, and R. Sorkin, astro-ph/0209227; S.
  Capozziello, S. Carloni, and A. Troisi, astro-ph/0303041; S.
  Carroll, V. Duvvuri, M. Trodden, and M. Turner, astro-ph/0306438.
  
\bibitem{SN1} A.~G. Riess, {\it et al.}  [Supernova Search Team
  Collaboration], Astron.\ J.\ {\bf 116}, 1009 (1998)
  [arXiv:astro-ph/9805201].
  
\bibitem{SN2} S. Perlmutter, {\it et al.}  [Supernova Cosmology
  Project Collaboration], Astrophys.\ J.\ {\bf 517}, 565 (1999)
  [arXiv:astro-ph/9812133].

  
\bibitem{mbs} I. Maor, R. Brustein, and P.J. Steinhardt, Phys. Rev.
  Lett.  {\bf 86}, 6 (2001); I. Maor, R. Brustein, J. McMahon, and
  P.J.  Steinhardt, Phys. Rev. D {\bf 65}, 123003 (2002).

\bibitem{barger} V. Barger and D. Marfatia, Phys. Lett. {\bf B498}, 67
(2001).

\bibitem{Wang01a} Y. Wang, and P. Garnavich, ApJ, 552, 445 (2001)
  
\bibitem{Wang01b} Y. Wang, and G. Lovelace, ApJ, 562, L115 (2001)
 
\bibitem{Tegmark02} M. Tegmark, Phys. Rev. D66, 103507 (2002)
 
\bibitem{Daly03} R.A. Daly, S.G. \& Djorgovski ApJ, 597, 9 (2003)
 
\bibitem{Phillips93} M.M. Phillips, ApJ, 413, L105 (1993)

\bibitem{Riess95} A.G. Riess, W.H. Press,  and R.P. Kirshner,  ApJ,
  438, L17 (1995)
  
\bibitem{omega1} C.B. Netterfield {\it et al}, Ap. J. {\bf 571}, 604
  (2002); R. Stompor {\it et al}, Ap. J. Lett. {\bf 561}, 7 (2001);
  N.W. Halverson {\it et al}, Ap. J. {\bf 568}, 38 (2002).
  
\bibitem{Weinberg72} S. Weinberg, ``Gravitation and Cosmology'' (John
  Wiley \& Sons, New York, 1972)

\bibitem{sch} P. Schueker, {\it et al}, Astron. Astrophys., {\bf 402},
53 (2003)

\bibitem{melchiorri} A. Melchiorri, L. Mersini, C.J. Odman, and M.
  Trodden, Phys.Rev. {\bf D68} 043509 (2003)
  
\bibitem{Press} W. H. Press, S. A. Teukolsky, W. T. Vetterling, 
  B.P. Flannery, ``Numerical Recipes in Fortran 77 (The Art of
  Scientific Computing)'' (Cambridge University Press, 2nd ed., 1992)
  
\bibitem{Wang00a} Y. Wang, ApJ, 531, 676 (2000a)
  
\bibitem{Wang00b} Y. Wang, ApJ, 536, 531 (2000b)
  
\bibitem{Wang04} Y. Wang, P. Mukherjee, astro-ph/0312192, ApJ
  in press (2004)
  
\bibitem{Tarle02} G. Tarle, (for the SNAP Collaboration),
  astro-ph/0210041
  

\bibitem{hutturn}  D. Huterer and M.S. Turner,
Phys.Rev. {\bf D64}   123527 (2001)

\bibitem{Linder} E.V. Linder and D. Huterer, Phys. Rev. D 67, 081303
  (2003)
  
\end{references}
\end{document}